\documentclass{article}
\newcommand{\be}{\begin{equation}}
\newcommand{\ee}{\end{equation}}
\newcommand{\bea}{\begin{eqnarray}}
\newcommand{\eea}{\end{eqnarray}}
\newcommand{\nn}{\nonumber \\}
\newcommand{\de}{{\rm d}}
\newcommand{\w}[1]{{\bf #1}}
\newcommand{\dens}{\rho_{\rm d}}
\newcommand{\Lap}{\triangle}
\newcommand{\frOm}{\frac 1{\Omega_{d-1}}}
\begin{document}

\centerline{\huge \bf Floating Bodies of Equilibrium}
\medskip

\centerline{\huge \bf at Density 1/2}
\medskip

\centerline{\huge \bf in Arbitrary Dimensions}

\vspace{1cm}

\begin{center}
\bf Franz Wegner, Institut f\"ur Theoretische Physik \\
Ruprecht-Karls-Universit\"at Heidelberg \\
Philosophenweg 19, D-69120 Heidelberg \\
Email: wegner@tphys.uni-heidelberg.de
\end{center}
\vspace{1cm}

\paragraph{Abstract} Bodies of density one half (of the fluid in which they are
immersed) that can float in all orientations are investigated.
It is shown that expansions starting from and deforming the (hyper)sphere are
possible in arbitrary dimensions and allow for a large manifold of solutions:
One may either (i) expand $r(\w n)+r(-\w n)$ in powers of
a given difference $r(\w u)-r(-\w u)$, ($r(\w n)$ denoting the distance from the
origin in direction $\w n$). Or (ii) the envelope of the water planes (for fixed
body and varying direction of gravitation) may be given.
Equivalently $r(\w n)$ can
be expanded in powers of the distance $h(\w u)$ of the water planes from the
origin perpendicular to $\w u$.

\section{Introduction and Results}

A long standing problem asked by Stanislaw Ulam in the Scottish
Book\cite{Scottish} (problem 19) is, whether a sphere is the only solid of
uniform
density which will float in water in any position. Such a solid is called a
{\it floating body of equilibrium}. It will be in indifferent equilibrium in all
orientations.

The simpler, two-dimensional, problem to find non circular cross-sections of a
long cylindrical log which floats without tending to rotate (the
axis of the log is assumed to be parallel to the water surface.) was solved for
relative density $\dens=1/2$ in 1938 by Auerbach \cite{Auerbach}.
He found a large class of solutions. This is in contrast to the solutions found
for $\dens\not=1/2$ in two dimensions\cite{WegnerI,WegnerII,WegnerIII} and for
central symmetric bodies in three dimensions\cite{Wegnerg}, where the variety
of shapes is much more restricted, provided one restricts to star-shaped
bodies.

Here it will be shown that also in dimensions $d>2$ there is a large variety of
bodies which can float in arbitrary orientation at $\dens=1/2$. More precisely:
Denote the distance from the origin to the boundary of the body by
$r(\w n)$, with the unit vector $\w n$ pointing into the direction.
Starting from the hypersphere $r(\w n)=r_0$ a deformation governed by an
expansion parameter $\epsilon$ will be considered
\be
r(\w n) = r_0 + \sum_{k=1}^{\infty} \epsilon^k r_k(\w n). \label{rexp}
\ee
Due to Archimedes' principle one half of the body is below the water, the other
half above. Thus the water plane cuts the body into two halves. Denote the
height of this plane above the origin by $h(\w n)$ with $\w n$ perpendicular to
the plane, then $h(-\w n)=-h(\w n)$. A similar expansion will be performed for
$h(\w n)$,
\be
h(\w n) = \sum_{k=1}^{\infty} \epsilon^k h_k(\w n). \label{hexp}
\ee
For practical reasons restriction will be made to solutions which obey
\bea
r_k(-\w n) &=& (-)^k r_k(\w n) \label{rcond} \\
h_{2k+1}(-\w n) &=& - h_{2k+1}(\w n), \quad h_{2k}(\w n)=0. \label{hcond}
\eea
In each order in $\epsilon$ there are two conditions on $r_k$ and $h_k$. One
condition guarantees that the volumes above and below the water level are equal
(V-condition, eq. \ref{Vhigh}), the other guarantees that the potential energy
does not depend on the orientation (Z-condition, eq. \ref{Zhigh}).

With the restrictions (\ref{rcond}, \ref{hcond}) one finds in odd orders
$\epsilon^{2k+1}$ that the Z-condition is identically fulfilled. The
V-condition yields that a linear combination of $r_{2k+1}$ and $h_{2k+1}$ is
determined by $r_n$ and $h_n$ with $n<2k+1$. In even orders $\epsilon^{2k}$ the
V-condition is identically fulfilled. The Z-condition determines $r_{2k}$.

This may be used in the following ways:

{\bf Given $r$.} If $r_1(\w n)$ is given, which obeys
$r_1(-\w n)=-r_1(\w n)$ and $r_{2k+1}=0$ for $k>0$,
then the conditions yield in odd order $h_k$ and in even order $r_k$.
If one chooses also $r_1=0$, then the $r_{2k}$ will also vanish (except $\w n$
independent contributions).
This is in agreement with the theorem by
Schneider\cite{Schneiderd,Schneidere} and Falconer\cite{Falconer}, also
referred to by Hensley in the Scottish book\cite{Scottish},
compare\cite{Wegnerg}: {\it For arbitrary dimension $d$ and
density $1/2$, if the body is star-shaped, symmetric, bounded and measurable,
then it differs from a ball by a set of measure 0.}

{\bf Given $h$.} If $h_1(\w n)$ is given, which obeys
$h_1(-\w n)=-h_1(\w n)$ and $h_k=0$ for $k>1$,
then the conditions yield $r_k$ in all orders.
Suppose the envelope of the water planes is given. The distance of the
planes tangent to this envelope yield $h(\w n)$. Thus to a given envelope one
may construct the surface of the floating body. If one rescales $r$ and $h$ by
a factor $1/\epsilon$, then one obtains
\be
r(\w n) = \frac{r_0}{\epsilon} + \sum_{k=1}^{\infty} \epsilon^{k-1} r_k(\w n),
\quad h(\w n) = h_1(\w n).
\ee
Thus for a given envelope one obtains a one-parameter manifold parametrized by
$\epsilon$. These envelopes have the property that there is exactly one tangent
hyperplane to them parallel to any given plane. As a consequence these
envelopes have cusps and are quite different from surfaces of convex bodies,
which have two tangent hyperplanes parallel to any given plane.
A one-parameter manifold for a given envelope was obtained in dimension $d=2$ by
Auerbach\cite{Auerbach}; see appendix \ref{appAuerbach}. From given
$h(\w n)$ one can determine the envelope (appendix \ref{appEnvelope}).

Basically the expansion scheme of ref. \cite{Wegnerg} will be used. In
subsection \ref{equ} the two conditions for equilibrium is given. Two
expansions are used: the expansion in $\epsilon$, which describes the deviation
from the hypersphere (\ref{epsexp}) and the expansion in ultraspherical
harmonics
(\ref{harmexp}). In subsection \ref{VZVZ} the expansion of $V$ and $Z$ in
ultraspherical harmonics is completed.
The calculation in first and second and order in $\epsilon$ is given in
subsections
\ref{first} and \ref{second}. The general discussion in higher orders is
presented in \ref{higher}. The paper is closed by a remark on reparametrization
with respect to the expansion in $\epsilon$ (\ref{rep}). Although the presented
arguments hold to arbitrary order in $\epsilon$, no
solution in closed form is presented nor even a proof of convergency for the
$\epsilon$-expansion.

\section{Equilibrium and Expansion}

\subsection{Equilibrium\label{equ}}

Due to Archimedes' principle the body obeys
\be
V_{\rm a} = V_{\rm b} = \frac 12 V_{\rm t}, \label{condV}
\ee
where $V_{\rm a}$ is the volume above, $V_{\rm b}$ the volume below the
water plane and $V_{\rm t}$ the total volume. Condition (\ref{condV}) determines
the height of the water plane.
 
The potential energy $\cal V$ of the body is given by
\bea
{\cal V} &=& \frac 12 m g (z_{\rm a}-z_{\rm b})
=\frac 12 \rho g V_{\rm t} (z_{\rm a}-z_{\rm b}) \nn
&=& \rho g (Z_{\rm a} - Z_{\rm b})
=2\rho g (Z_{\rm a}-\frac 12 Z), \
\eea
with $m$ and $\rho$ mass and density of the body, $g$ the
gravitational acceleration, $z_{\rm a}$ and $z_{\rm b}$ the $z$-coordinates of
the part of the body above and below the water plane, (the $z$-coordinate
measured
in direction of $\w n$, $Z_{\rm a}$, $Z_{\rm b}$, and $Z_{\rm t}$ the integral
of $z$
over the volume above, below the water plane and over the total volume. In order
that the body can float freely in all orientations the potential energy has to
be independent of the orientation. Thus $Z_{\rm a}-\frac 12 Z_{\rm t}$ has to 
be constant.

\subsection{Expansion\label{epsexp}}

In the following a unit vector $\w u$ in $d$ dimensions is decomposed into the
component upward and a vector $\w u'$ in the $d-1$-dimensional space parallel
to the water plane
\be
\w u = \cos\theta\, \w n+ \sin\theta\, \w u'
\ee
Then the element $\de\Omega_d$ of the solid angle in $d$ dimensions can be
written
\be
\de\Omega_{\w u} = \de\Omega_{\w u'}\, \de\theta\, (\sin\theta)^{d-2}.
\ee
The angle $\Theta$, at which the water plane intersects the surface of the body
as a function of $\w u'$ has to be determined from
\be
h = r\cos\Theta.
\ee
Expanding
\be
\Theta=\frac{\pi}2 +\delta\Theta, \quad
r=r_0+\delta r, \quad
\delta r(\Theta) = \sum_{j=0}^{\infty} \frac{(\delta\Theta)^j}{j!} \delta
r^{(j)}
\ee
(the $\delta r^{(j)}$ are the coefficients of a Taylor expansion of $\delta r$
in $\delta\Theta$) one obtains the equation
\be
h = -(r_0 +\sum_{j=0}^{\infty} \frac{(\delta\Theta)^j}{j!} \delta r^{(j)})
\sin\delta\Theta, \label{htheta}
\ee
which has to be solved for $\delta\Theta$ for arbitrary $\w u'$, which is not
indicated explicitly. This eq. is invariant against the transformation
$h\rightarrow -h$, $\delta\Theta\rightarrow -\delta\Theta$,
$\delta r^{(j)}\rightarrow (-)^j\delta r^{(j)}$, thus $\delta\Theta$ obeys the
symmetry relation
\bea
\delta\Theta_+ &=& -\delta\Theta_-, \label{Thetasym} \\
\delta\Theta_+ := \delta\Theta(h,\{\delta r^{(j)}\}), &&
\delta\Theta_- := \delta\Theta(-h,\{(-)^j\delta r^{(j)}\}).
\eea
The volume $V_{\rm a}$ and $Z_{\rm a}$ are obtained by realizing that the
volume,
which is a hypersegment, can be divided into a hypersector (first integral) and
a hypercone (second integral)
\bea
V_{\rm a} &=& \frac 1d \int\de\Omega_{\w u'} \int_0^{\Theta(\w u')} \de\theta\,
r^d(\theta,\w u')(\sin\theta)^{d-2} \nn
&-& \frac h{(d-1)d} \int\de\Omega_{\w u'} 
\Big(r(\Theta(\w u'),\w u') \sin(\Theta(\w u'))\Big)^{d-1}, \\
Z_{\rm a} &=& \frac 1{d+1} \int\de\Omega_{\w u'} \int_0^{\Theta(\w u')}
\de\theta\, r^{d+1}(\theta,\w u')\cos\theta(\sin\theta)^{d-2} \nn
&-& \frac{h^2}{(d-1)(d+1)} \int\de\Omega_{\w u'}
\Big(r(\Theta(\w u'),\w u') \sin(\Theta(\w u'))\Big)^{d-1}
\eea
It is practical to split the $\theta$-integration into one from 0 to $\pi/2$
and a second one from $\pi/2$ to $\pi/2+\delta\Theta(\w u')$.
Then the expressions for $V_{\rm a}$ and $Z_{\rm a}$ are given by
\bea
V_{\rm a} = V_{\rm a}^{(1)} + V_{\rm a}^{(2)}, &&
Z_{\rm a} = Z_{\rm a}^{(1)} + Z_{\rm a}^{(2)}, \\
V_{\rm a}^{(1)} &=& \frac 1d \int\de\Omega_{\w u'} \int_0^{\pi/2} \de\theta\,
r^d(\theta,\w u')(\sin\theta)^{d-2}, \\
Z_{\rm a}^{(1)} &=& \frac 1{d+1} \int\de\Omega_{\w u'} \int_0^{\pi/2}
\de\theta\,
r^{d+1}(\theta,\w u')\cos\theta(\sin\theta)^{d-2}, \\
V_{\rm a}^{(2)} = \int\de\Omega_{\w u'}\, g, &&
Z_{\rm a}^{(2)} =\int\de\Omega_{\w u'}\, g^z \label{VZ2}
\eea
with
\bea
g(h,\{\delta r^{(j)}\})
&=& I_g - \frac h{(d-1)d} 
\Big(r(\frac{\pi}2+\delta\Theta_+(\w u')) \cos(\delta\Theta_+(\w
u')\Big)^{d-1}, \label{intg2} \\
I_g(h,\{\delta r^{(j)}\})
&=& \frac 1d \int_0^{\delta\Theta_+} \de\theta'\,
r^d(\frac{\pi}2+\theta',\{\delta r^{(j)}\})
\cos^{d-2}\theta', \\
g^z(h,\{\delta r^{(j)}\}
&=& I_{g^z}
- \frac{h^2}{(d-1)(d+1)}
\Big(r(\frac{\pi}2+\delta\Theta_+(\w u')) \cos(\delta\Theta_+(\w
u')\Big)^{d-1}, \label{intgz2} \\
I_{g^z}(h,\{\delta r^{(j)}\})
&=& \frac{-1}{d+1} \int_0^{\delta\Theta_+} \de\theta'\,
r^{d+1}(\frac{\pi}2+\theta',\{\delta r^{(j)}\})
\cos^{d-2}\theta'\sin\theta'.
\eea
Replace
\be
r(\frac{\pi}2+\theta',\{\delta r^{(j)}\}) =
r(\frac{\pi}2-\theta',\{(-)^j\delta r^{(j)}\})
\ee
and in a second step $\theta'$ by $-\theta'$. Then $I_g$ reads
\be
I_g= -\frac 1d \int_0^{-\delta\Theta_+} \de\theta'
r^d(\frac{\pi}2+\theta',\{(-)^j\delta r^{(j)}\}) \cos^{d-2}\theta'.
\ee
Insertion of (\ref{Thetasym}) yields for the integral
\be
I_g(h,\{\delta r^{(j)}\}) = -I_g(-h,\{(-)^j\delta r^{(j)}\}).
\ee
Similarly one shows
\be
I_{g^z}(h,\{\delta r^{(j)}\})
= I_{g^z}(-h,\{(-)^j\delta r^{(j)}\}).
\ee
Since
\be
r(\frac{\pi}2+\delta\Theta_+,\{\delta r^{(j)}\})
=r(\frac{\pi}2-\delta\Theta_+,\{(-)^j\delta r^{(k)}\})
=r(\frac{\pi}2+\delta\Theta_-,\{(-)^j\delta r^{(j)}\})
\ee
one observes that the second terms in (\ref{intg2}) and (\ref{intgz2}) obey the
same symmetry relation so that the relations
\bea
g(h,\{\delta r^{(j)}\}) &=& -g(-h,\{(-)^k\delta r^{(j)}\}), \label{gsym2} \\
g^z(h,\{\delta r^{(j)}\}) &=& g^z(-h,\{(-)^j\delta r^{(j)}\}) \label{gzsym2}
\eea
hold. In performing these calculations $h$ depends on $\w n$, whereas $\delta r$
depends on $\w u'$. Since $\delta r$ is expanded around $\theta=\pi/2$ the
$\delta r^{(j)}$ are to be taken at $\w u'$ perpendicular to $\w n$. Thus $g$
and $g^z$ depend on
\be
h=h(\w n), \quad \delta r^{(j)} = \delta r^{(j)}(\w u').
\ee

\subsection{Expansion in ultraspherical harmonics\label{harmexp}}

The quantities under the $\Omega_{\w u'}$ integrals are expanded in
ultraspherical harmonics
\be
f(\w u) = f(\cos\theta\w n+\sin\theta\w u')
= \sum_{l=0}^{\infty} f_{;l}(\w u),
\ee
where the ultraspherical harmonics are eigenfunctions of the Laplacian on the
unit sphere
\be
\Lap_{\w u} f_{;l}(\w u) = -l(l+d-2 ) f_{;l}(\w u).
\ee
To distinguish this expansion from that in $\epsilon$ the index $l$ is always
preceeded by a semicolon. For given $d$ and $l$ there are
\be
\#^d_l = (2l+d-2) \frac{(l+d-3)!}{(d-2)!l!}
\ee
linearly independent ultraspherical harmonics.

The integral over $\Omega_{\w u'}$ yields a function
depending only on $\cos\theta$. The integral over $f_{;l}$ yields
\be
\int\de\Omega_{\w u'} f_{;l}(\w u)
= \Omega_{d-1} f_{;l}(\w n) \frac{C_l^{(d/2-1)}(\cos\theta)}{C_l^{(d/2-1)}(1)}
\ee
with the ultraspherical (Gegenbauer) polynomial $C_l^{(d/2-1)}(x)$, which is the
only function $f_{;l}$ independent of $\w u'$. The factor in front is obtained,
since the integral is the factor $\Omega_{d-1}$ times the average over $\w u'$,
which for $\theta\rightarrow 0$ approaches $f_{;l}(\w n)$. In appendix
\ref{Formula} a few formula for theses polynomials are listed. See also
\cite{Abramowitz,Bateman}.

In evaluating the integrals for $V^{(2)}_{\rm a}$ and $Z^{(2)}_{\rm a}$
(\ref{VZ2}) one has to integrate over $\w u'$ at $\theta=\frac{\pi}2$.
Thus
\bea
\int\de\Omega_{\w u'} f_{;l}(\w u')
&=&\Omega_{d-1} \gamma^{(d)}_l f_{;l}(\w n), \\
\gamma^{(d)}_l &=& \frac{C_l^{(d/2-1)}(0)}{C_l^{(d/2-1)}(1)}
=\left\{\begin{array}{cc} (-)^l
\frac{\Gamma(\frac{l+1}2)\Gamma(\frac{d-1}2)}{\sqrt{\pi}\Gamma(\frac{l+d-1}2)}
& {\rm even}\,\, l \\ 0 & {\rm odd}\,\, l \end{array}\right. .
\eea
These quantities are listed in appendix \ref{Dim23} for $d=2$ and 3.

\subsection{$V_{\rm t}$, $Z_{\rm t}$, $V_{\rm a}^{(1)}$, and $Z_{\rm a}^{(1)}$
\label{VZVZ}}

The total volume is given by
\bea
V_{\rm t} &=& \frac 1d \int\de\Omega_{\w u'} \int_0^{\pi} \de\theta\, 
r^d(\theta,\w u')\sin^{d-2}\theta \nn
&=& \frac 1d \Omega_{d-1} \sum_l (r^d)_{;l}(\w n) \int_0^{\pi} \de\theta\,
\frac{C^{(d/2-1)}_l(\cos\theta)}{C^{(d/2-1)}_l(1)} \sin^{d-2}\theta.
\eea
The integral over $\theta$ vanishes for all $l$ except $l=0$,
\be
V_{\rm t} = \frac 1d \Omega_d\, (r^d)_{;0}(\w n).
\ee
There is only one ultraspherical harmonic $C^{(d/2-1)}_0(x)=1$ for $l=0$.
$(r^d)_{;0}$ is a scalar and does not depend on $\w n$.

$Z_{\rm t}$ of the total volume is given by
\bea
Z_{\rm t} &=& \frac 1{d+1} \int\de\Omega_{\w u'} \int_0^{\pi} \de\theta\, 
r^{d+1}(\theta,\w u')\cos\theta\sin^{d-2}\theta \nn
&=& \frac 1{d+1} \Omega_{d-1} \sum_l (r^{d+1})_{;l}(\w n) \int_0^{\pi}
\de\theta\,
\frac{C^{(d/2-1)}_l(\cos\theta)}{C^{(d/2-1)}_l(1)} \cos\theta\sin^{d-2}\theta.
\eea
The integral over $\theta$ vanishes for all $l$ except $l=1$,
\be
Z_{\rm t} = \frac 1{d(d+1)} \Omega_d\, (r^{d+1})_{;1}(\w n).
\ee
There are $d$ ultraspherical harmonics for $l=1$. They transform under 
rotations like the components of a vector. $(r^{d+1})_{;1}(\w n)$ is the 
projection of volume times the vector to the centre of gravity onto $\w n$.

One obtains
\be
\frOm (V_{\rm a}^{(1)} - \frac 12 V_{\rm t})
= \frac 1d \sum_l I^{(v)}_l(r^d)_{;l}(\w n)
\ee
with
\be
I^{(v)}_l = \frac{(d-2) C^{(d/2)}_{l-1}(0)}{l(l+d-2)C^{(d/2-1)}_l(1)}
=\left\{\begin{array}{cc} 0 & {\rm even}\,\, l \\
 (-)^{(l-1)/2}
 \frac{\Gamma(\frac l2)\Gamma(\frac{d-1}2)}{2\sqrt{\pi}\Gamma(\frac{l+d}2)}
  & {\rm odd}\,\, l \end{array} \right. . \label{Iv}
\ee
and
\be
\frOm (Z_{\rm a}^{(1)} - \frac 12 Z_{\rm t})
= \frac 1{d+1} \sum_l I^{(z)}_l (r^{d+1})_{;l}(\w n)
\ee
with
\bea
I^{(z)}_l &=& \frac{d-2}{(2l+d-2)C^{(d/2-1)}_l(1)}
\left(\frac{C^{(d/2)}_l(0)}{l+d-1}+\frac{C^{(d/2)}_{l-2}(0)}{l-1}\right) \nn
&=& \left\{\begin{array}{cc} (-)^{(l-2)/2}
\frac{\Gamma(\frac{l-1}2)\Gamma(\frac{d-1}2)}{4\sqrt{\pi}\Gamma(\frac{l+d+1}2)}
& {\rm even}\,\, l \\ 0 & {\rm odd}\,\, l \end{array}\right. . \label{Iz}
\eea

\subsection{First Order in $\epsilon$\label{first}}

Eq. (\ref{htheta}) yields in first order in $\epsilon$
\be
\delta\Theta_1 = -\frac{h_1}{r_0}.
\ee
One obtains
\bea
\frOm (Z^{(1)}_{\rm a,1}-\frac 12 Z_{\rm t,1}) &=& r_0^d \sum_l I^{(z)}_l
r_{1;l}(\w n), \\
I_{g^z,1}=0, && g^z_1=0, \qquad Z_{\rm a,1}^{(2)}=0.
\eea
Thus
\be
I^{(z)}_l r_{1;l}(\w n) = 0 \mbox{ for } l\not= 0 \label{Iz1}
\ee
has to be fulfilled. Since $I^{(z)}_l$ vanishes for odd $l$ one can choose
$r_{1;l}$ arbitrarily for odd $l$, whereas $r_{1;l}=0$ for even $l$ with the
exception of $l=0$. $r_{1;0}\not=0$ would change the radius. For simplicities'
sake $r_{1;0}=0$ is chosen.

Further one obtains
\bea
\frOm (V^{(1)}_{\rm a,1}-\frac 12 V_{\rm t,1})
&=& r_0^{d-1} \sum_l I^{(v)}_l r_{1;l}(\w n), \\
I_{g,1}=-\frac{h_1}d r_0^{d-1}, && g_1=-\frac{h_1}{d-1} r_0^{d-1}, \\
\frOm V^{(2)}_{\rm a,1} &=& -\frac{1}{d-1} h_{1;l} r_0^{d-1},
\eea
which yields
\be
h_{1;l} = (d-1)I^{(v)}_l r_{1;l}. \label{Iv1}
\ee
Thus one obtains contributions $h_{1;l}$ proportional to $r_{1;l}$ for odd $l$,
whereas $h_{1;l}$ vanishes for even $l$, since both $r_{1;l}$ and $I^{(v)}_l$
vanish for even $l$.

\subsection{Second Order in $\epsilon$\label{second}}

In second order in $\epsilon$ one obtains
\bea
\delta\Theta_2 &=& -\frac{h_2}{r_0} + \frac{r_1h_1}{r_0^2}, \\
\frOm (Z^{(1)}_{\rm a,2}-\frac 12 Z_{\rm t,2})
&=& r_0^{d-1} \sum_l I^{(z)}_l (r_0r_2+\frac d2 r_1^2)_{;l}, \\
I_{g^z,2} &=& -\frac 1{2(d+1)}h_1^2r_0^{d-1}, \\
g^z_2 &=&-\frac 1{2(d-1)}h_1^2r_0^{d-1}, \\
\frOm Z_{\rm a,2}^{(2)} &=& -\frac 1{2(d-1)}h_1^2r_0^{d-1}.
\eea
This yields the equation
\be
I^{(z)}_l (r_0r_2+\frac d2 r_1^2)_{;l}
-\frac 1{2(d-1)}(h_1^2)_{;l} = 0.
\ee
For odd $l$ one obtains $I^{(z)}_l=0$, $(r_1^2)_{;l}=0$, $(h_1^2)_{;l}=0$. Thus
$r_{2;l}$ can be chosen arbitrarily, but in conformity with (\ref{rcond}), the
choice $r_{2;l}=0$ is made for odd $l$. $r_{2;l}$ is determined for even $l$ by
\be
r_{2;l} = -\frac{d(r_1^2)_{;l}}{2r_0} + \frac 1{2(d-1)I^{(z)}_l}(h_1^2)_{;l}.
\label{Iz2}
\ee
One obtains for $V$ 
\bea
\frOm (V^{(1)}_{\rm a,2}-\frac 12 V_{\rm t,2}) &=& r_0^{d-2} \sum_l I^{(v)}_l
(r_0r_2+\frac{d-1}2 r_1^2)_{;l}, \\
I_{g,2} &=& -\frac 1{2d}r_0^{d-1}h_2(\w n)
-\frac{2d-1}{2d} r_1(\w u')h_1(\w n)r_0^{d-2}, \\
g_2 &=& -\frac{d+1}{2d(d-1)}r_0^{d-1}h_2(\w n) -\frac{2d+1}{2d}r_0^{d-2}r_1(\w
u')h_1(\w n), \\
\frOm V^{(2)}_{\rm a,2} &=& \frac{d+1}{2d(d-1)}r_0^{d-1}h_2(\w n) \nn
&-& \sum_l\frac{2d+1}{2d} \gamma^{(d)}_l r_0^{d-2}(r_1)_{;l}(\w n) h_1(\w n)
\eea
$V^{(1)}_{\rm a,2}-\frac 12 V_{\rm t,2}$ vanishes, since
$(r_0r_2+\frac{d-1}2 r_1^2)_{;l}$ vanishes for odd $l$ and $I^{(v)}_l$ for even
$l$.
The second term in the expression for $V^{(2)}_{\rm a,2}$ vanishes, since
$(r_1)_{;l}$ vanishes for even $l$ and $C^{(d/2-1)}_l(0)$ for odd $l$.
Thus the condition
$(V^{(1)}_{\rm a,2}+V^{(2)}_{\rm a,2}-\frac 12 V_{\rm t,2})_{;l}=0$
reduces to $h_2(\w n)=0$.

Thus it has been shown that (\ref{rexp}) and (\ref{hexp}) hold up to second
order in $\epsilon$. Next it will be shown by complete induction that they
hold in arbitrary order.

\subsection{Higher Orders in $\epsilon$\label{higher}}

$g_k$ and $g^z_k$ are polynomials in $h_q$ and $\delta r_p^{(j)}$. They are
linear combinations of terms
\be
\prod_m h_{q_m} \prod_m \delta r_{p_m}^{(j_m)}
\ee
In order $\epsilon^k$ contribute terms with
\be
k=\sum q_m + \sum p_m.
\ee
Except for linear terms in the polynomials $g_k$ and $g_k^z$, which have to be
determined, use of (\ref{rcond}) yields
\be
\prod_m \delta r_{p_m}^{(j_m)}(-\w u')
= (-)^{\sum_m j_m + \sum_m p_m} \prod_m \delta r_{p_m}^{(j_m)}(\w u').
\ee
Thus only terms with even $\sum_m j_m + \sum_m p_m$ contribute to
$V^{(2)}_{{\rm a},k}$ and $Z^{(2)}_{{\rm a},k}$, eq. (\ref{VZ2}). From eqs.
(\ref{gsym2}) and (\ref{gzsym2}) one deduces $\sum q_m + \sum j_m$ is odd for
$g_k$ and even for $g^z_k$. Due to (\ref{hcond}) only
$h_q$ with odd $q$ can differ from zero. Thus all non-linear terms (n.l.t.) to
$V^{(2)}_k$ vanish for even $k$ and such contributions to $Z^{(2)}_k$ vanish
for odd $k$.

Next consider the contributions $V^{(1)}_{{\rm a},k}-\frac 12 V_{{\rm t},k}$
and $Z^{(1)}_{{\rm a},k}-\frac 12 Z_{{\rm t},k}$. Eq. (\ref{rexp}) yields
\be
(r^d)_k(-\w u) = (-)^k (r^d)_k(\w u), \quad 
(r^{d+1})_k(-\w u) = (-)^k (r^{d+1})_k(\w u).
\ee
Since $f_{;l}(-\w u) = (-)^l f_{;l}(\w u)$ there are only contributions
$(r^d)_{k;l}$ and $(r^{d+1})_{k;l}$ with even $k-l$.

Extracting the terms linear in $r_k$ and $h_k$ one obtains
\bea
\frOm (Z_{\rm a}-\frac 12 Z_{\rm t})_k &=& r_0^d \sum_l I_l^{(z)} r_{k;l}
\label{Zhigh}
+ \mbox{ n.l.t.}, \\
\frOm (V_{\rm a}-\frac 12 V_{\rm t})_k &=& r_0^{d-1} \sum_l
(I_l^{(v)} r_{k;l} -\frac 1{d-1} h_{k;l}) + \mbox{ n.l.t.}. \label{Vhigh}
\eea

Consider odd $k$. $Z^{(2)}_{{\rm a},k}$ vanishes, the n.l.t. in
$Z^{(1)}_{{\rm a},k}$ vanish, too, since $I^{(z)}_l=0$ for odd $l$ and all
n.l.t. contribute to odd $l$. One is free to choose $r_{k;l}$ with odd $l$.
One obtains $h_{k;l}$ from (\ref{Vhigh}). Since all n.l.t. contribute to odd
$l$, only $h_{k;l}$ with odd $l$ can differ from zero.

Consider even $k$. Then all n.l.t. from $Z$ contribute to even $l$. They
determine $r_{k;l}$ with even $l$. For odd $l$ the n.l.t. vanish and since
$I^{(z)}_l=0$ for odd $l$ one could choose non-zero $r_{k;l}$. However, choose
$r_{k;l}=0$ for odd $l$ and even $k$ in agreement with (\ref{rcond}). 
All n.l.t. to $V_{{\rm a},k}$ vanish.
Moreover $I^{(v)}_lr_{k;l}=0$, since the first factor vanishes for even $l$,
the second for odd $l$. Thus $h_k=0$ for even $k$. This completes the proof.

\subsection{Reparametrization\label{rep}}

In each order $\epsilon^k$ it is possible to choose $r_{k;l}$ for odd $l$
freely. We
restricted ourselves to do this for odd $k$ only. Basically it is sufficient to
do this for $k=1$ only by reparametrizing
\be
r'_{1;l} = \sum_{k=1}^{\infty} \epsilon^{2k} r_{2k+1;l}
\ee
and $r'_{2k+1;l}=0$ for $k>0$, since it will yield the same result $r(\w n)$.

If one requires that the centre of gravity is located at the origin, then one
has to adjust $r_{k;1}$ in all odd orders $k$. Then in first order it has to be
$r_{1;1}=0$. Thus the restriction (\ref{rcond}) does not restrict the manifold
of solutions.

It is not necessary to introduce an expansion in $\epsilon$. Alternatively one
expands $r(\w n)+r(-\w n)-2r_0$ in powers of $r(\w u)-r(-\w u)$.

\begin{appendix}

\section{Appendices}

\subsection{Auerbach's Solution\label{appAuerbach}}

Basically Auerbach\cite{Auerbach} starts from the envelope of the water line.
The length $2\ell$ of the water line is constant. The line touches the envelope
in the middle. Thus the envelope may be represented by
\be
x_{\rm e}(\phi) = x_0 + \int_0^{\phi} \de\phi' s(\phi') \cos(\phi'), \quad
y_{\rm e}(\phi) = y_0 + \int_0^{\phi} \de\phi' s(\phi') \sin(\phi').
\ee
The water line and thus the point, at which the envelope touches the water line
has to be
the same after $\phi$ has increased by $\pi$. Thus 
\be
x_{\rm e}(\pi) = x_{\rm e}(0), \quad y_{\rm e}(\pi) = y_{\rm e}(0),
\quad s(\pi+\phi) = -s(\phi)
\ee
is required, which yields
\be
x_{\rm e}(\phi+\pi) = x_{\rm e}(\phi), \quad
y_{\rm e}(\phi+\pi) = y_{\rm e}(\phi).
\ee
Then the boundary of the body is given by
\be
x(\phi) = x_{\rm e}(\phi) + \ell \cos\phi, \quad
y(\phi) = y_{\rm e}(\phi) + \ell \sin\phi.
\ee
The scheme presented here is less elegant than the procedure used by
Auer\-bach, however the present scheme applies to general dimensions. It
would be desirable to have an elegant procedure which does not refer to a
perturbation expansion for the construction of such bodies in $d>2$ dimensions.

\subsection{Envelope from $h(\w n)$\label{appEnvelope}}

For given
$h(\w n)$ one can determine the corresponding point $\w r_{\rm e}(\w n)$ on the
envelope. It obeys
\be
(\w n+\delta\w n)\cdot \w r_{\rm e}(\w n) = h(\w n+\delta\w n)
\ee
to first order in $\delta\w n$, which is orthogonal to $\w n$. Expanding both
sides
\be
\w n\cdot\w r_{\rm e} + \delta\w n\cdot\w r_{\rm e}
=h(\w n)+\delta\w n\cdot\nabla h(\w n)
\ee
yields
\be
\w r_{\rm e}(\w n) = \w n h(\w n) + \nabla h(\w n).
\ee

\subsection{Ultraspherical (Gegenbauer) Polynomials\label{Formula}}

A few relations on ultraspherical (Gegenbauer) polynomials are given.
Rodrigues' formula
\be
C^{\alpha}_n(x) = (-)^n \frac{\Gamma(\alpha+1/2)\Gamma(n+2\alpha)}
{2^n n!\Gamma(2\alpha)\Gamma(n+\alpha+1/2)}
(1-x^2)^{-\alpha+1/2} \frac{\de^n}{\de x^n} (1-x^2)^{n+\alpha-1/2} \label{DefC}
\ee
Recurrence relation
\be
2(n+\alpha)xC_n^{\alpha}(x)
= (n+1) C_{n+1}^{\alpha}(x) +(n+2\alpha-1) C_{n-1}^{\alpha}(x).
\ee
Special values
\be
C^{\alpha}_n(\pm 1) = (\pm 1)^n \left( {n+2\alpha-1\atop n}\right).
\ee
\be
C^{\alpha}_{2m}(0) = (-)^m \left({m+\alpha-1 \atop m}\right), \quad
C^{\alpha}_{2m+1}(0) = 0.
\ee
Integrals
\bea
&& \int \de\theta (\sin\theta)^{d-2} C_l^{(d/2-1)}(\cos\theta)\de\theta \nn
&=&\frac{d-2}{l(l+d-2)} (\sin\theta)^{d-1} C_{l-1}^{(d/2)}(\cos\theta), 
\quad l>0 \\
&& \int \de\theta \cos\theta(\sin\theta)^{d-2}
C_l^{(d/2-1)}(\cos\theta)\de\theta \nn
&=& \left\{ \begin{array}{ll}
\frac{d-2}{2l+d-2} (\sin\theta)^{d-1}
\left(\frac 1{l+d-1}C_l^{(d/2)}(\cos\theta)
+ \frac 1{l-1} C_{l-2}^{(d/2)}(\cos\theta)\right), & l>1, \\
\frac 1{d-1} (\sin\theta)^{d-1}, & l=0
\end{array} \right.
\eea

The full solid angle is obtained from
\be
\Omega_d = \Omega_{d-1} \int_0^{\pi}\de\theta \sin^{d-2}\theta
= \Omega_{d-1} \frac{\sqrt{\pi}\Gamma(\frac{d-1}2)}{\Gamma(\frac d2)}
\ee
with $\Omega_1=2$ for the two directions in one dimension, which yields
\be
\Omega_d = \frac{2\pi^{d/2}}{\Gamma(\frac d2)}.
\ee
The duplication formula for the $\Gamma$-function reads
\be
\Gamma(2z) = \frac 1{\sqrt{\pi}} 2^{2z-1} \Gamma(z) \Gamma(z+\frac 12).
\ee

\subsection{Dimensions two and three\label{Dim23}}

\subsubsection{The limit $d\rightarrow 2$}

The normalization of $C^{(d/2-1)}_l$, (\ref{DefC}) is not convenient for the
limit $d\rightarrow 2$, since $C$ approaches 0 in this limit for $l\not=0$.
One obtains the limit
\be
\lim_{d\rightarrow 2} \frac{C^{(d/2-1)}_l(\cos\theta)}{C^{(d/2-1)}_l(1)}
= T_l(\cos\theta) = \cos(l\theta),
\ee
where $T_l$ denotes the Chebyshev polynomials. That this normalization is not
convenient, can also be seen from the factors $(d-2)$ in (\ref{Iv}, \ref{Iz}).
However, one finds in this limit\\
\be
\begin{array}{c|cc}
 & {\rm even}\,\, l & {\rm odd}\,\, l \\ \hline
\gamma^{(2)}_l & (-)^l & 0 \\
I^{(v)}_l & 0 & (-)^{(l-1)/2} \frac 1l \\
I^{(z)}_l & (-)^{(l-2)/2} \frac 1{l^2-1} & 0
\end{array}
\ee

\subsubsection{Three Dimensions}

In three dimensions one obtains
\be
C^{1/2}_l(\cos\theta) = P_l(\cos\theta)
\ee
with the Legendre polynomials $P_l$.
One obtains
\be
\begin{array}{c|cc}
 & {\rm even}\,\, l & {\rm odd}\,\, l \\ \hline
\gamma^{(3)}_l & (-)^l\frac{l!}{2^l(\frac l2)!^2} & 0 \\
I^{(v)}_l & 0 & (-)^{(l-1)/2} \frac{(l-1)!}{2^l(\frac{l+1}2)!(\frac{l-1}2)!} \\
I^{(z)}_l & (-)^{(l-2)/2} \frac{l!}{2^l(l-1)(l+2)(\frac l2)!^2} & 0
\end{array}
\ee

\end{appendix}

\end{document}